\documentclass[twocolumn,aps,prd,preprintnumbers,nofootinbib]{revtex4-2}
\usepackage{graphicx}
\usepackage{amsmath}
\usepackage{amssymb}
\usepackage{amsfonts}
\usepackage{hyperref}
\usepackage{url}
\usepackage{color}
\usepackage{bm}
\usepackage[dvipsnames]{xcolor}
\usepackage{comment}

\newcommand{\be}{\begin{equation}}
\newcommand{\ee}{\end{equation}}
\newcommand{\ba}{\begin{eqnarray}}
\newcommand{\ea}{\end{eqnarray}}

\renewcommand{\[}{\begin{equation}}
\renewcommand{\]}{\end{equation}}

\def\be{\begin{equation}}
\def\ee{\end{equation}}
\def\bea{\begin{eqnarray}}
\def\eea{\end{eqnarray}}
\def\eqi{\begin{equation}}
\def\eqf{\end{equation}}
\def\eqia{\begin{eqnarray}}
\def\eqfa{\end{eqnarray}}
\def\lcdm{$\Lambda$CDM}
\newcommand{\HH}{\mathcal{H}}
\newcommand{\hJ}{\hat{J}}

\def\lcdm{$\Lambda$CDM}

\usepackage{array}
\usepackage{ulem}

\def\dsbt#1{\textcolor{brown}{#1}}

\makeatother

\begin{document}

\preprint{IFT-UAM/CSIC-24-132}

\title{Probing \lcdm\ through the Weyl potential and machine learning forecasts}

\author{Rub\'{e}n Arjona}
\email{rarjona@ucm.es}

\author{Savvas Nesseris}
\email{savvas.nesseris@csic.es}

\affiliation{Instituto de F\'isica Te\'orica UAM-CSIC, Universidad Auton\'oma de Madrid,
Cantoblanco, 28049 Madrid, Spain}

\author{Isaac Tutusaus}
\email{isaac.tutusaus@irap.omp.eu}

\affiliation{Institut de Recherche en Astrophysique et Planétologie (IRAP), Université de Toulouse,
CNRS, UPS, CNES, 14 Av. Edouard Belin, F-31400 Toulouse, France}

\author{Daniel Sobral Blanco}
\email{daniel.sobralblanco@unige.ch}

\author{Camille Bonvin}
\email{camille.bonvin@unige.ch}

\affiliation{Département de Physique Théorique and Center for Astroparticle Physics,
Université de Genève, Quai E. Ansermet 24, CH-1211 Genève 4, Switzerland}

\date{\today}

\begin{abstract}
For years, the cosmological constant $\Lambda$ and cold dark matter (CDM) model ($\Lambda\text{CDM}$) has stood as a cornerstone in modern cosmology and serves as the predominant theoretical framework for current and forthcoming surveys. However, the latest results shown by the Dark Energy Spectroscopic Instrument (DESI), along other cosmological data, show hints in favor of an evolving dark energy. Given the elusive nature of dark energy and the imperative to circumvent model bias, we introduce a novel null test, derived from Noether's theorem, that uses measurements of the Weyl potential (the sum of the spatial and temporal distortion) at different redshifts. In order to assess the consistency of the concordance model we quantify the precision of this null test through the reconstruction of mock catalogs based on \lcdm\ using forthcoming survey data, employing Genetic Algorithms, a machine learning technique. Our analysis indicates that with forthcoming LSST-like and DESI-like mock data our consistency test will be able to rule out several cosmological models at around 4$\sigma$ and help to check for tensions in the data.
\end{abstract}

\maketitle

\section{Introduction\label{sec:intro}}
Recent observations over the last couple of decades have led to overwhelming support in favor of the cosmological constant ($\Lambda$) and Cold Dark Matter (CDM) model, known as the \lcdm\ model \cite{Aghanim:2018eyx}. It
significantly outperforms alternative models, as it has been noted via Bayesian analyses of cosmological measurements \cite{Heavens:2017hkr}. However, the dynamics of dark energy (DE) remain inadequately scrutinized. Moreover, discrepancies have emerged within the framework of \lcdm, manifesting at different levels of statistical significance (most notably, the ``Hubble tension \cite{di2021snowmass2021}", the ``$\sigma_8$ tension \cite{di2021cosmology}" and an evolving DE in light of the observations from the Dark Energy Spectroscopic Instrument (DESI) \cite{adame2024desi}). Should these discrepancies persist beyond potential unaccounted-for systematic errors, they may signify novel physics beyond \lcdm.

In this context, several consistency tests and analyses have been proposed \cite{Clarkson:2007pz,Shafieloo:2009hi,Mortsell:2011yk,Sapone:2014nna,Rasanen:2014mca,LHuillier:2016mtc,Denissenya:2018zcv,Park:2017xbl,Cao:2021ldv,Khadka:2020tlm,Cao:2021ldv,DiDio:2016ykq,Vardanyan:2009ft}. Despite the potential presence of unaccounted systematic errors contributing to the observed discrepancies, there remains a plausible avenue for new physics, potentially in the guise of modified gravity (MG) or alternative DE models. Notably, the \lcdm\ framework carries inherent limitations, in particular 
in its primary constituents—CDM and DE which have yet to be directly detected in laboratory settings and remain poorly understood \cite{Bertone:2016nfn,Weinberg:1988cp,Carroll:2000fy}. This suggests that \lcdm\ may serve as an approximation to a deeper, currently inaccessible theory \cite{DiValentino:2020vhf}. The proliferation of MG and DE models complicates observational interpretations, with cosmological parameter outcomes, such as the Universe's matter density fraction $\Omega_\textrm{m,0}$, contingent upon the specific model employed. Recent efforts have concentrated on establishing a unified framework to encompass these diverse models, such as the Effective Field Theory (EFT) \cite{Gubitosi:2012hu,Hu:2013twa} or the Effective Fluid Approach (EFA) \cite{Arjona:2018jhh,Arjona:2019rfn,Arjona:2020gtm,Cardona:2020ama}.

To mitigate the biases inherent in selecting a theoretically predefined model, non-parametric reconstruction techniques and model-independent approaches offer promising avenues  \cite{Nesseris:2010ep}. Within this context, machine learning (ML) algorithms have emerged as innovative tools for extracting information in a theory-agnostic manner \cite{Ntampaka:2019udw}. These methods are particularly adept at identifying potential tensions arising from unaccounted systematics or offering insights into novel physics. Their key advantage lies in their ability to signal deviations at various redshifts, indicating the breakdown of underlying assumptions \cite{Marra:2017pst}. Null tests have found extensive application in testing the \lcdm\ model \cite{Sahni:2008xx,Zunckel:2008ti,Nesseris:2010ep,arjona2022testing,arjona2021complementary}, interacting DE models \cite{vonMarttens:2018bvz}, the growth-rate data \cite{Nesseris:2014mfa,Marra:2017pst,Benisty:2020kdt}, the cosmic curvature \cite{Clarkson:2007pz, Yahya:2013xma,Cai:2015pia,Benisty:2020otr,Li:2014yza}, the equivalence principle~\cite{Bonvin:2020cxp}, the distance duality relation \cite{Martinelli:2020hud} and also to probe the scale-independence of the growth of structure in the linear regime \cite{Franco:2019wbj}. A primary advantage of null tests lies in their ability to signal any deviations from expected values across various redshifts, indicating the breakdown of underlying assumptions \cite{Marra:2017pst}. Unlike traditional model-fitting approaches, null tests do not rely on a specific model; instead, they function as pass-fail assessments. These tests complement traditional model-fitting analyses by offering the capability to potentially reject multiple models simultaneously. They achieve this by simultaneously scrutinizing the fundamental assumptions of numerous models.

The main purpose of this work is to present a new consistency test for the \lcdm\ model using measurements of the Weyl potential (i.e.\ the sum of the spatial and temporal distortion of the metric) and of the Hubble parameter $H(z)$. Assuming general relativity (GR) applied to a perturbed Friedmann-Lema\^itre-Robertson-Walker metric, we can derive a
second order differential equation to describe the evolution of the Weyl potential. Then, assuming a flat \lcdm\ model we apply Noether's theorem to construct the aforementioned null test.  We apply the Genetic Algorithms (GA), a stochastic minimization and symbolic regression algorithm, to reconstruct the Weyl potential and the $H(z)$ function from
LSST-like and DESI-like mock data respectively. Our null test is quite generic and has to be valid at all redshifts. We show that with surveys like the LSST and DESI, the Weyl potential and $H(z)$ data will be able to discriminate a wide range of MG theories from $\Lambda$CDM.  Notably, GA's non-parametric nature minimizes assumptions regarding the underlying cosmology, thus circumventing potential biases.

The outline of our paper is as follows: in Sec.~\ref{sec:theory} we introduce our theoretical framework. In Sec.~\ref{sec:nulltest} we set out our \lcdm\ consistency test via Noether's theorem and in Sec.~\ref{sec:Weyl} we present the parametrization considered for the Weyl potential. In 
Sec.~\ref{sec:GA} we describe the data used and our machine learning forecasts. Then, in Sec.~\ref{sec:results} we present our results and in Sec.~\ref{sec:conclusions} we summarize our conclusions.

\section{Theoretical Framework\label{sec:theory}}
At sufficiently large scales $\gtrsim 100 \mathrm{Mpc}$ the Universe can be described by a flat Friedmann-Lema\^itre-Robertson-Walker (FLRW) metric. In order to investigate the perturbations within different cosmological frameworks, we consider the perturbed FLRW metric, which in the conformal Newtonian gauge takes the form:
\be
ds^2=a(\tau)^2\left\{-\left[1+2\Psi(\vec{x},\tau)\right]d\tau^2+\left[1-2\Phi(\vec{x},\tau)\right]d\vec{x}^2\right\},
\label{eq:FRWpert}
\ee
where $\tau$ is the conformal time defined via $d\tau\equiv dt/a(t)$, $t$ is the cosmic time and we are following the notation of Ref.~\cite{Ma:1995ey}.\footnote{In more detail, our conventions are: (-+++) for the metric signature, the Riemann and Ricci tensors are given by $V_{b;cd}-V_{b;dc}=V_a R^a_{bcd}$ and $R_{ab}=R^s_{asb}$, while the Einstein equations are $G_{\mu\nu}=+\kappa T_{\mu\nu}$ for $\kappa=\frac{8\pi G_N}{c^4}$ and $G_N$ is the bare Newton's constant. In what follows we will set the speed of light $c=1$.} At this stage we can assume an ideal fluid with an energy momentum tensor
\be
T^\mu_{\nu}=P\,\delta^\mu_{\nu}+(\rho+P)\,U^\mu U_\nu,\label{eq:enten}
\ee
where $\rho$, $P$ are the fluid density and pressure, while $U^\mu\equiv\frac{dx^\mu}{\sqrt{-ds^2}}$ is its velocity four-vector given to first order by $U^\mu=\frac{1}{a(\tau)}\left(1-\Psi,\vec{u}\right)$, which satisfies $U^\mu U_\mu=-1$. Furthermore, $\vec{u}=\dot{\vec{x}}$, where $\dot{f}\equiv\frac{df}{d\tau}$, and the elements of the energy momentum tensor to first order of perturbations are given by:
\bea
T^0_0&=&-(\bar{\rho}+\delta \rho),\\
T^0_i&=&(\bar{\rho}+\bar{P})u_i,\\
T^i_j&=& (\bar{P}+\delta P)\delta^i_j+\Sigma^i_j, \label{eq:effectTmn}
\eea
where $\bar{\rho},\bar{P}$ are defined on the background and are functions of time only, while the perturbations $\delta \rho, \delta P$ are functions of $(\vec{x},\tau)$ and $\Sigma^i_j\equiv T^i_j-\delta^i_j T^k_k/3$ is an anisotropic stress tensor. Given GR as our framework, we find that the perturbed Einstein equations in the conformal Newtonian gauge take the form of \cite{Ma:1995ey}:
\be
k^2\Phi+3\frac{\dot{a}}{a}\left(\dot{\Phi}+\frac{\dot{a}}{a}\Psi\right) = 4 \pi G_N a^2 \delta T^0_0, \label{eq:phiprimeeq}
\ee
\be
k^2\left(\dot{\Phi}+\frac{\dot{a}}{a}\Psi\right) = 4 \pi G_N a^2 (\bar{\rho}+\bar{P})\theta,\label{eq:phiprimeeq1}
\ee
\be
\ddot{\Phi}+\frac{\dot{a}}{a}(\dot{\Psi}+2\dot{\Phi})+\left(2\frac{\ddot{a}}{a}-
 \frac{\dot{a}^2}{a^2}\right)\Psi+\frac{k^2}{3}(\Phi-\Psi)
=\frac{4\pi}{3}G_N a^2\delta T^i_i,\label{eq:phiprimeeq2}
\ee
\be
k^2(\Phi-\Psi) = 12\pi G_N a^2 (\bar{\rho}+\bar{P})\sigma \label{eq:anisoeq},
\ee
where we have defined the velocity $\theta\equiv ik^ju_j$ and the anisotropic stress  $(\bar{\rho}+\bar{P})\sigma\equiv-(\hat{k}_i\hat{k}_j-\frac13 \delta_{ij})\Sigma^{ij}$. 

Since our aim is to build a consistency test of $\Lambda$CDM, we set the anisotropic stress to zero at late time, leading to $\Phi=\Psi$. Combining Eqs.~(\ref{eq:phiprimeeq}) and (\ref{eq:phiprimeeq2})  we can get an equation for $\Psi$ alone: 
\begin{align}\label{eq:potential}
    \ddot{\Psi} + & 3 \mathcal{H}\left[1-\frac{1}{3}\frac{\delta T^i_i}{\delta T^0_0}\right]\dot{\Psi} \\
    &\qquad+\left[\mathcal{H}^2\left(1-\frac{\delta T^i_i}{\delta T^0_0}\right)+2\dot{\mathcal{H}}-k^2\frac{1}{3}\frac{\delta T^i_i}{\delta T^0_0}\right] \Psi=0, \nonumber
\end{align}
where we have introduced the conformal Hubble factor $\mathcal{H} = \dot{a}/a$. Eq.~\eqref{eq:potential} can be further simplified by noting that at late time (when radiation and neutrinos are negligible) $\frac{\delta T^i_i}{\delta T^0_0}=0$, since CDM and baryons are pressure-less and the cosmological constant has no perturbations.

Then Eq.~(\ref{eq:potential}) reads:
\begin{equation}\label{eq:potential1}
\ddot{\Psi}+3 \mathcal{H}\dot{\Psi}+\left(2\dot{\mathcal{H}}+\mathcal{H}^2\right)\Psi=0.
\end{equation}
We can also express the above equation in terms of the scale factor as
\begin{equation}\label{eq:potential2}
\Psi^{\prime \prime}+\left(\frac{4}{a}+\frac{\mathcal{H}^{\prime}}{\mathcal{H}}\right) \Psi^{\prime}+\left(\frac{1}{a^2}+\frac{2 \mathcal{H}^{\prime}}{a \mathcal{H}}\right) \Psi=0,
\end{equation}
where the prime is the derivative with respect to the scale factor. From Eq.~\eqref{eq:potential2} we see that in \lcdm\ the evolution of the potential at late times (and ignoring radiation and neutrinos) does not depend on the scale $k$.

\section{Lagrangian formalism and null test}\label{sec:nulltest}
In seeking a null test that encompasses the evolution of the potential, we will employ the Lagrangian formalism. Initially, we aim to derive a Lagrangian for Eq.~(\ref{eq:potential}) and, with the help of 
Noether’s theorem, to find an associated conserved quantity. If we assume that the Lagrangian can be written as $\mathcal{L}=\mathcal{L}\left(a, \Psi(a), \Psi^{\prime}(a)\right)$, where $a$, $\Psi(a)$ and $\Psi^{\prime}(a)$ are the “time”, the generalized position and velocity variables of the system, respectively, then the
Euler-Lagrange equations are:
\begin{equation}\label{eq:E-L}
\frac{\partial \mathcal{L}}{\partial \Psi}-\frac{d}{d a} \frac{\partial \mathcal{L}}{\partial \Psi^{\prime}}=0.
\end{equation}
So, let us assume a Lagrangian of the form
\ba
\mathcal{L} &=&T-V, \\
T &=&\frac{1}{2} f_{1}(a, \mathcal{H}(a)) \Psi^{\prime}(a)^{2}, \\
V &=&\frac{1}{2} f_{2}(a, \mathcal{H}(a)) \Psi(a)^{2},
\ea
where 
$f_1$ and $f_2$ are arbitrary functions that need to be determined so that the resulting equation after implementing the Euler-Lagrange Eq.~(\ref{eq:E-L}) is exactly Eq.~(\ref{eq:potential}). Therefore, after making the substitution we are able to get the two functions $f_1$ and $f_2$ and consequently to build the Lagrangian $\mathcal{L}$ of the system:
\begin{equation}\label{eq:lagrange}
\mathcal{L}=\frac{1}{2} a^4 \mathcal{H} \Psi^{\prime}(a)^2-\frac{1}{2}\left(a^2 \mathcal{H}+2 a^3 \mathcal{H}^{\prime}\right) \Psi^2(a).
\end{equation}
 With the Lagrangian in hand, we can utilize Noether’s theorem to identify a conserved quantity, which will subsequently be applied to formulate the null test. So, if we have an infinitesimal transformation $\mathbf{X}$ with a generator
\ba 
\mathbf{X} &=&\alpha(\Psi) \frac{\partial}{\partial \Psi}+\frac{d \alpha(\Psi)}{d a} \frac{\partial}{\partial \Psi^{\prime}}, \\ \frac{d \alpha(\Psi)}{d a} & \equiv & \frac{\partial \alpha}{\partial \Psi} \Psi^{\prime}(a)=\alpha^{\prime}(a),
\ea
such that for the Lie derivative of the Lagrangian we have $L_{X} \mathcal{L}=0$, then
\begin{equation}\label{eq:sigma}
\Sigma=\alpha(a) \frac{\partial \mathcal{L}}{\partial \Psi^{\prime}}
\end{equation}
is a constant of “motion” for the Lagrangian of Eq.~(\ref{eq:lagrange}). From Eq.~(\ref{eq:sigma}) we get that
\begin{equation}
\Sigma=a^4\mathcal{H}\alpha(a)\Psi^{\prime}(a),
\end{equation}
while from the Lie derivative we also obtain:
\begin{equation}
\alpha(a)=c\,e^{\int_1^a \frac{\Psi(x)\left(\mathcal{H}(x)+2 x \mathcal{H}^{\prime}(x)\right)}{\mathcal{H}(x) x^2 \Psi^{\prime}(x)} d x},
\end{equation}
where $c$ is an integration constant. Then the constant is given by
\begin{equation}
\Sigma=a^4 \mathcal{H}(a) \Psi^{\prime}(a)\,e^{\int_1^a \frac{\Psi(x)\left(\mathcal{H}(x)+2 x \mathcal{H}^{\prime}(x)\right)}{\mathcal{H}(x) x^2 \Psi^{\prime}(x)} d x},
\end{equation}
where we have redefined $\Sigma$ to absorb $c$. If we normalize the above equation we have the following null test that must be 1 for all values of the scale factor $a$,
\begin{equation}\label{eq:nulltest}
\mathcal{O}(a)=a^4 \frac{\mathcal{H}(a)}{\mathcal{H}(1)} \frac{\Psi^{\prime}(a)}{\Psi^{\prime}(1)}\,e^{\int_1^a \frac{\Psi(x)\left(\mathcal{H}(x)+2 x \mathcal{H}^{\prime}(x)\right)}{\mathcal{H}(x) x^2 \Psi^{\prime}(x)} d x},
\end{equation}
which is the main result of our paper. It can also be written in terms of the redshift $z$ through $a=\frac{1}{1+z}$. Eq.~(\ref{eq:nulltest}) is applicable solely within the framework of a perturbed FLRW universe featuring a cosmological constant, devoid of perturbations. Any deviations from this standard scenario would necessitate alterations to Eq.~(\ref{eq:nulltest}). For instance, if we consider a DE model other than a cosmological constant, wherein DE exhibits clustering behavior, these perturbations would affect the fluid's sound speed. If we discard the assumption of a DE energy 
component entirely, and instead entertain MG theories, Eq.~(\ref{eq:nulltest}) itself must be modified, resulting in a distinct evolution of the potential. In summary, any deviation from unity in $\mathcal{O}(a)$ may signal one of four different scenarios:
\begin{itemize}
    \item A deviation from the perturbed FLRW metric.
    \item Non-zero DE  
    perturbations.
    \item A deviation from GR, i.e.\ MG 
    models.
    \item Tension between the $\mathcal{H}(z)$ and $\Psi(z)$ data.
\end{itemize}

\section{Weyl potential}\label{sec:Weyl}
The time distortion $\Psi$ entering into Eq.~\eqref{eq:nulltest} has never been measured in the linear regime. It impacts galaxy clustering through the effect of gravitational redshift~\cite{Bonvin:2011bg,Sobral-Blanco:2022oel} and will be detectable with the coming generation of galaxy surveys~\cite{Bonvin:2013ogt,Bonvin:2015kuc}. However, since in $\Lambda$CDM the two gravitational potentials are the same, we can use instead measurements of the Weyl potential $\Psi_W=(\Phi+\Psi) / 2 = \Psi$, which has recently been measured from gravitational lensing~\cite{Tutusaus:2023aux}. 

More precisely, combining galaxy-galaxy lensing and galaxy clustering allows one to measure the function $\hJ(z)$, which governs the evolution of the Weyl potential at late time through the relation (see~\cite{Tutusaus:2022cab,Tutusaus:2023aux} for more detail) 
\begin{align}
\Psi_W(k,z)= \left[\frac{\HH(z)}{\HH(z_*)} \right]^2 \sqrt{\frac{B(k,z)}{B(k,z_*)}}\hJ(z)\frac{\Psi_W(k,z_*)}{\sigma_8(z_*)}\, . 
\label{eq:hatJ}
\end{align}
Here $z_*$ is a reference redshift, that we choose to be well in the matter era, and $\sigma_8(z_*)$ denotes the amplitude of density fluctuations at that redshift. $B(k,z)$ is a boost factor, encoding the non-linear evolution of matter density perturbations at small scales. The boost is introduced in standard lensing 
analyses 
to properly account for non-linearities that affect the lensing correlation function at small angular separation. However, since in the $\Lambda$CDM model $\hJ$ is scale-independent, once it has been measured over a specific range of scales (those accessible to the lensing survey), it can be used at any scales, i.e.\ for any value of $k$. In this work, we concentrate on large scales, in the linear regime, and we use $\hJ$ in this regime where $B(k,z)/B(k,z_*)=1$.

In this regime, we see from Eq.~\eqref{eq:hatJ}, that the Weyl potential $\Psi_W$ (which enters the null test) and the function $\hJ$ (which can be directly 
measured from the data) are related by the Hubble parameter $\HH(z)$. Hence the null test can be rewritten in terms of the observable $\hJ$. More precisely, inserting Eq.~\eqref{eq:hatJ} into Eq.~\eqref{eq:nulltest}, in the regime where $B(k,z)/B(k,z_*)=1$, we can rewrite the two ratios that enter in the null test as
\be
\label{eq:Psi_J}
\frac{\Psi(a)}{\Psi^{\prime}(a)}=\frac{\mathcal{H}(a)\hat{J}(a)}{2\mathcal{H}^{\prime}(a)\hat{J}(a)+\mathcal{H}(a)\hat{J}^{\prime}(a)},
\ee
and
\be
\label{eq:Psip_J}
\frac{\Psi^{\prime}(a)}{\Psi^{\prime}(1)}=\frac{\mathcal{H}(a)\left(2\mathcal{H}^{\prime}(a)\hat{J}(a)+\mathcal{H}(a)\hat{J}^{\prime}(a)\right)}{2\mathcal{H}^{\prime}(1)\hat{J}(1)+\mathcal{H}(1)\hat{J}^{\prime}(1)}.
\ee
To build the null test we need therefore measurements of $\HH, \hJ$ and their derivatives.

\section{Reconstructions\label{sec:GA}}
Even though $\hJ$ has already been measured with Dark Energy Survey (DES) Year 3 data~\citep{Tutusaus:2023aux}, we use mock data from the next generation of galaxy surveys to perform our test. The reason is that the null test depends on the evolution of $\hJ$ with redshift, and DES data only allowed us to measure $\hJ$ at 4 redshifts, which would not provide a faithful reconstruction of $\hJ'$. We consider therefore mock data from a survey like LSST. Similarly for $\HH(z)$ we consider mock data from a future BAO survey. In the following we detail the specificity of the mock data employed and the machine learning methodology utilized for reconstructing the null test, specifically employing GA. 

\subsection{Mock data \label{sec:data}}
Our mock Hubble rate data ($H(z)$) is derived from anticipated enhancements to the DESI \cite{Aghamousa:2016zmz}. Designed to explore the Universe's expansion rate and large-scale structure (LSS), DESI complements other forthcoming Baryon Acoustic Oscillation (BAO) surveys, broadening the range of redshifts under examination \cite{Martinelli:2020hud}.

Commencing operations in late 2019, the DESI survey aims to collect optical spectra from tens of millions of galaxies and quasars up to redshifts nearing $z\sim 4$. This extensive reach enables comprehensive cosmological analyses involving BAO and redshift-space distortion. Our forecasted data spans the redshift interval $z\in [0.05,3.55]$, with precision contingent upon the targeted galactic populations. Specifically, Bright Galaxy Sample (BGS) are observed across $z\in [0.05,0.45]$ in five equally spaced bins, while luminous red galaxies (LRGs), emission line galaxies (ELGs), and Lyman-alpha forest quasars (QSOs) are scrutinized across distinct redshift ranges: $z\in [0.65,1.85]$ with 13 bins, and $z\in [1.96,3.55]$ with 11 bins, respectively \cite{Aghamousa:2016zmz}. This differentiation in redshift distributions among targets reflects varied selection methods necessary to accumulate sizable spectroscopic samples from photometric data, as detailed in Section 3 of \cite{Aghamousa:2016zmz}.

We follow Ref.~\cite{Aghamousa:2016zmz}, distributing $H(z)$ uniformly within the interval $z\in[0.1,3.55]$, divided into 20 equidistant bins with a step size of $dz=0.2$. Each value of $H(z_i)$ is estimated theoretically from different cosmological models, to which we add noise drawn from a normal distribution with error equal to $0.5\%$ of the value of $H(z)$. We assume these measurements to be uncorrelated. 

Concerning the forecast data for $\hJ(z)$, we consider the full data release of LSST~\citep{LSST:2008ijt}. This corresponds to 18\,000 square degrees over 10 years of observations in multiple optical bands. This survey will enable us to accurately measure photometric redshifts and shear measurements for about 27 galaxies per arcmin$^2$, leading to exquisite measurements of galaxy-galaxy lensing.

In this work we follow~\cite{Tutusaus:2022cab} and consider the public LSST specifications provided in the \textsc{CosmoSIS} framework~\citep{Zuntz:2014csq}. These correspond to 5 tomographic bins for the source galaxies used for weak lensing and 10 tomographic bins for the lens galaxies used for galaxy clustering. We assume the same galaxy number density of 27 galaxies per arcmin$^2$ for both the lens and source populations. We further account for astrophysical systematic effects like galaxy bias or intrinsic alignments. For the former, we consider a linear galaxy bias model with a fiducial galaxy bias $b=2$ in each bin. We let these amplitudes vary independently and we marginalize over them. Concerning the intrinsic alignment modeling, we consider the nonlinear alignment model~\citep{Bridle:2007ft,2012MNRAS.424.1647K} with a fiducial amplitude set to 1 but allowed to vary. We study two different scenarios in our analysis: an optimistic case where we consider multipoles between $\ell=20$ and $\ell = 2627$, as provided in \textsc{CosmoSIS}, and a pessimistic case where our maximum multipole is set to $\ell = 750$. We finally consider a total ellipticity dispersion of $\sigma_{\epsilon}=0.3$.

We note that we do not account for systematic uncertainties in this work, like a shear calibration bias or residuals in the estimation of the different galaxy distributions. Furthermore, we neglect some subdominant contributions to the model like redshift-space distortions or magnification, for simplicity. Applications of the null test presented in this work to real observations will need to account for these higher-order contributions in the model and systematic effects. However, the forecast methodology considered in this analysis is enough to show the relevance of the null test with future observations.

\subsection{Genetic algorithms}
In this section, we briefly summarize the GA and how they are incorporated in our analytical framework. Widely adopted in cosmology, the GAs have
demonstrated efficacy in diverse reconstructions across a broad spectrum of data. For comprehensive discussions, readers are referred to various references \cite{Bogdanos:2009ib,Nesseris:2010ep, Nesseris:2012tt,Nesseris:2013bia,Sapone:2014nna,Arjona:2020doi,Arjona:2020kco,Arjona:2019fwb,Arjona:2021hmg,Arjona:2020skf,Arjona:2020axn,aizpuru2021machine}. Beyond cosmology, GA find 
application in particle physics \cite{Abel:2018ekz,Allanach:2004my,Akrami:2009hp}, astronomy, and astrophysics \cite{wahde2001determination,Rajpaul:2012wu,Ho:2019zap}. Further exploration of symbolic regression methods in physics and cosmology can be found in \cite{Udrescu:2019mnk,Setyawati:2019xzw,vaddireddy2019feature,Liao:2019qoc,Belgacem:2019zzu,Li:2019kdj,Bernardini:2019bmd,Gomez-Valent:2019lny}.

The GA, a specific type of machine learning method, excel 
in unsupervised regression tasks, adept at conducting non-parametric reconstructions by identifying analytical functions that accurately describe the data. Emulating biological evolution, GA operate 
via the principles of natural selection, facilitated by genetic operations like mutation and crossover. In essence, candidate functions evolve over time through stochastic processes such as crossover—combining candidate functions to produce new ones—and mutation—randomly altering candidate functions. This iterative process iterates thousands of times to ensure convergence, with different random seeds exploring distinct regions of the functional space.

As a stochastic approach, GA assign 
probabilities to populations of functions giving rise to offspring based on their fitness to the data, typically measured by a $\chi^2$ statistic in our analysis, indicating the agreement of each function with the data. For our simulated data, we assume Gaussian likelihoods, thus employing the $\chi^2$ statistic within the GA approach. This probability mechanism, coupled with fitness evaluation, exerts evolutionary pressure favoring the fittest functions in each population, effectively driving the fit toward the minimum within a few generations.

In our analysis we reconstruct the Hubble rate $H(z)$ and the function  $\hat{J}(z)$ from the mock data created, and the procedure unfolds as follows. First, our predefined grammar was formed on the following  functions: exp, log, polynomials etc. and a set of operations $+,-,\times,\div$, see Table \ref{tab:grammars} for the complete list.

\begin{table}[!t]
\caption{The grammars used in the GA analysis. Other complex forms are automatically produced by the mutation and crossover operations as described in the text.\label{tab:grammars}}
\begin{centering}
\begin{tabular}{cc}
 Grammar type & Functions \\ \hline
Polynomials & $c$, $x$, $1+x$ \\
Fractions & $\frac{x}{1+x}$\\
Trigonometric & $\sin(x)$, $\cos(x)$, $\tan(x)$\\
Exponentials & $e^x$, $x^x$, $(1+x)^{1+x}$ \\
Logarithms & $\log(x)$, $\log(1+x)$
\end{tabular}
\par
\end{centering}
\end{table}

For our $H(z)$ reconstruction, we set the prior that $H(z=0)=H_0$, while for the reconstruction of the function $\hat{J}(z)$, we assume that the Universe at early times went through a phase of matter domination $\left( z\simeq 100\right)$, during which the Weyl potential was constant and $\hJ(a) \simeq a$ at high redshifts. These priors stem from physical considerations, reflecting the observed value of the Hubble parameter today and the Hubble law. Notably, we refrain from presuming the curvature of the Universe or adopting any specific modified gravity or DE model. Additionally, to prevent overfitting or spurious reconstructions, we enforce continuity and differentiability for all functions reconstructed by the GA, 
ensuring no singularities occur within the redshift range covered by the data.

Following the construction of the initial population, each member's fitness is evaluated using a $\chi^2$ statistic, with the $H(z)$ and $\hat{J}(z)$ data points serving as direct input. Subsequently, via a tournament selection process detailed in Ref.~\cite{Bogdanos:2009ib}, the best-fitting functions from each generation are chosen, and stochastic operations—crossover and mutation—are applied. To ensure convergence, the GA process iterates thousands of times with various random seeds, thoroughly exploring the functional space. Ultimately, the output comprises sets of functions describing the Hubble rate $H(z)$ and the function $\hJ(z)$ respectively. Note that the Hubble parameter in conformal time $\mathcal{H}(a)$, which enters the null test in~\eqref{eq:nulltest},~\eqref{eq:Psi_J} and~\eqref{eq:Psip_J} is related to $H(a)$ through $\mathcal{H}(a)=aH(a)$, while their derivatives with respect to $a$ obey $\mathcal{H}'(a)=H(a)+aH'(a)$.

The error estimates for the reconstructed function are derived using the path integral approach, originally introduced in Refs.~\cite{Nesseris:2012tt,Nesseris:2013bia}. This method involves analytically estimating the error of the reconstructed quantity by computing a path integral over all conceivable functions surrounding the best-fit GA
solution contributing to the likelihood. Notably, this error estimation technique remains robust irrespective of the correlation status of the data points. Extensive scrutiny and validation against a bootstrap Monte Carlo method have been conducted, as detailed in Ref.~\cite{Nesseris:2012tt}. To elaborate, upon obtaining a reconstructed function $f(x)$ from the GA, the path integral approach outlined in Ref.~\cite{Nesseris:2012tt} provides the $1\sigma$ error $\delta f(x)$. Extensive comparisons have been performed, confirming the appropriateness of the error analysis in our work.

The necessity for utilizing GA arises from the limitations of traditional inference approaches, like Markov chain Monte Carlo simulations, which demand a specific expansion history model, e.g., \lcdm, to fit the data. Such model-dependent approaches risk biasing the results or overlooking critical data features. The principal advantage of GA lies in its ability to determine the best-fit model from the data, allowing for a theory-agnostic approach without presuming a specific DE model.


\begin{figure*}[!thb]
\centering
\includegraphics[width = 0.48\textwidth]{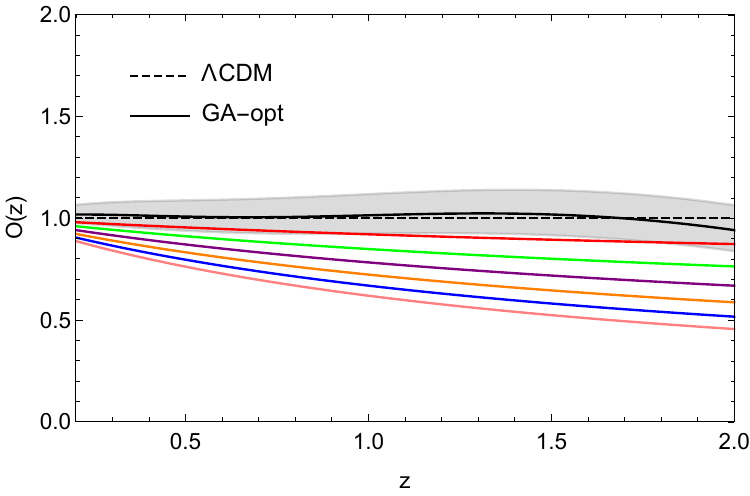}
\includegraphics[width = 0.48\textwidth]{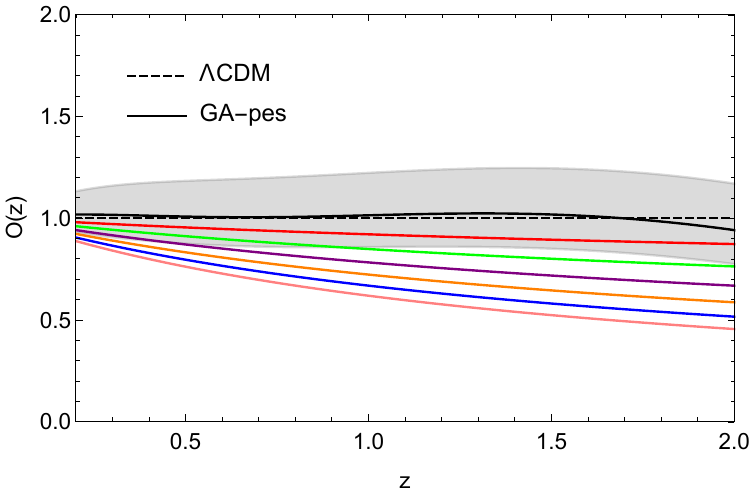}\caption{Reconstructions of the $\mathcal{O}(z)$ test of Eq.~\eqref{eq:nulltest}. The black solid line corresponds to the GA best fit, the gray shaded region corresponds to the $68.3\%$ confidence regions and the black dashed line to the \lcdm\ prediction. On the left panel we consider the optimistic case and on the right panel the pessimistic case for the measurements of $\hJ$. In both panels we show also $\mathcal{O}(z)$ in different models beyond $\Lambda$CDM, namely considering different values for the equation of state $w$ but neglecting DE perturbations, i.e.\ setting $\frac{\delta T^i_i}{\delta T^0_0}=0$. Specifically, the values are: $w$=\{-0.95 red, -0.9 green, -0.85 purple, -0.8 orange, -0.75 blue, -0.7 pink\}.  \label{fig:Oz_w}}

\end{figure*}

\begin{figure*}[!thb]
\centering
\includegraphics[width = 0.48\textwidth]{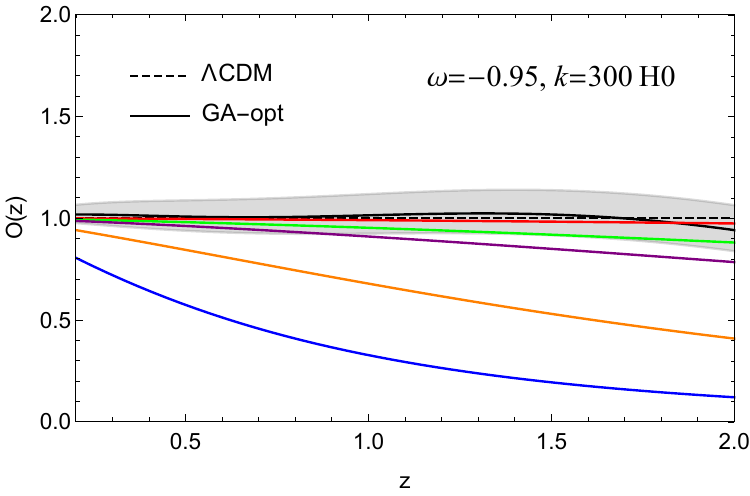}
\includegraphics[width = 0.48\textwidth]{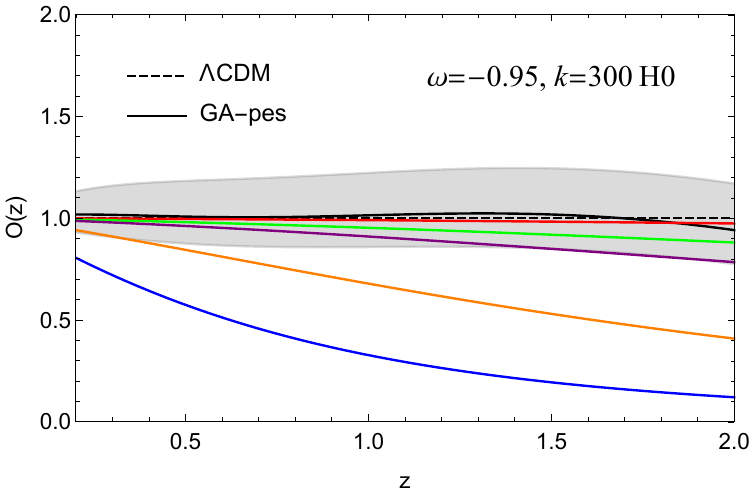}\caption{Reconstructions of the $\mathcal{O}(z)$ test of Eq.~\eqref{eq:nulltest}. The black solid line corresponds to the GA best fit, the gray shaded region corresponds to the $68.3\%$ confidence regions and the black dashed line to the \lcdm\ prediction. On the left panel we consider the optimistic case and on the right panel the pessimistic case for the measurements of $\hJ$. In both panels we show also $\mathcal{O}(z)$ in different models beyond $\Lambda$CDM, namely considering a fix value for the equation of state $w=-0.95$, and different values of the effective sound speed $c_{s,{\rm eff}}^2$. We show the results for the wave-number $k=300\,H_0\sim 0.1\,h\, \mathrm{Mpc}^{-1}$ (see Sec.\ref{sec:results} for a detailed explanation). Specifically, the values are: $c_{s,{\rm eff}}^2$=\{$1\cdot10^{-7}$ red, $5\cdot10^{-7}$ green, $1\cdot10^{-6}$ purple, $5\cdot10^{-6}$ orange, $1\cdot10^{-5}$ blue\}.   \label{fig:Oz_cs2}}
\end{figure*}

\begin{figure*}[!thb]
\centering
\includegraphics[width = 0.48\textwidth]{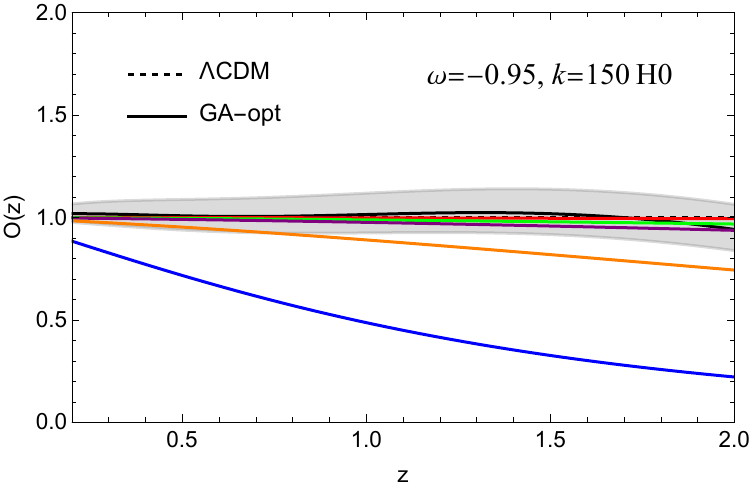}
\includegraphics[width = 0.48\textwidth]{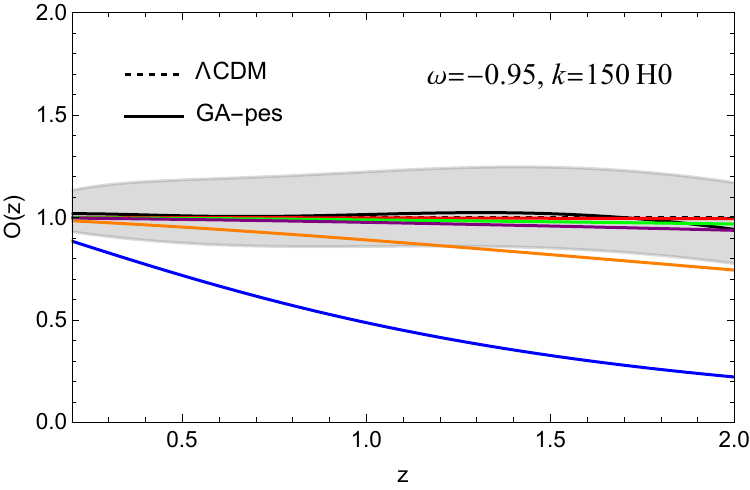}\caption{Reconstructions of the $\mathcal{O}(z)$ test of Eq.~\eqref{eq:nulltest}. The black solid line corresponds to the GA best fit, the gray shaded region corresponds to the $68.3\%$ confidence regions and the black dashed line to the \lcdm\ prediction. On the left panel we consider the optimistic case and on the right panel the pessimistic case for the measurements of $\hJ$. In both panels we show also $\mathcal{O}(z)$ in different models beyond $\Lambda$CDM, namely considering a fix value for the equation of state $w=-0.95$, and different values of the effective sound speed $c_{s,{\rm eff}}^2$. We show the results for the wave-number $k =150\,H_0\sim 0.05 h \,\mathrm{Mpc}^{-1}$ (see Sec.\ref{sec:results} for a detailed explanation). In specific the values are: $c_{s,{\rm eff}}^2$=\{$1\cdot10^{-7}$ red, $5\cdot10^{-7}$ green, $1\cdot10^{-6}$ purple, $5\cdot10^{-6}$ orange, $1\cdot10^{-5}$ blue\}. \label{fig:Oz_cs2_1}}
\end{figure*}

\section{Results \label{sec:results}}

In this section, we present our GA fits to the simulated data and the corresponding consistency test results obtained from our reconstructions of $H(z)$ and $\hat{J}(z)$. Our reconstructed functions start at $z=0.1$ which is the redshift of our first mock data point up to $z=2$. 

In Fig.~\ref{fig:Oz_w} we present the reconstruction of the $\mathcal{O}(z)$ test of Eq.~\eqref{eq:nulltest}. The black solid line corresponds to the GA best fit, the gray shaded region corresponds to the $68.3\%$ confidence regions and the black dashed line to the \lcdm\ prediction. Recall that the \lcdm\ curve is theoretical and so is precisely at $\mathcal{O}(z)=1$.  On the left panel we consider the optimistic case and on the right panel the pessimistic case, which correspond to fixing $\ell_{\rm max}=2627$ and $\ell_{\rm max}=750$ in the forecasts for $\hJ$, respectively. We see that in the optimistic case, $\mathcal{O}(z)$ can be reconstructed with a precision of $4.7-11.9\%$, which degrades to $11.2-21.8\%$ in the pessimistic case. This shows that forthcoming data from DESI and LSST will provide an efficient way of measuring precisely $\mathcal{O}(z)$ over a wide redshift range. Any deviation from unity would be a smoking gun for a deviation from the FLRW metric, an equation of state of DE different from minus 1, non zero DE perturbations, deviations from GR or a tension between the $H(z)$ and $\hat{J}(z)$ data. 

To assess the sensitivity of $\mathcal{O}(z)$ to deviations from \lcdm, we compute it in two alternative scenarios. First we consider the case where the equation of state of DE $w$ differs from minus one, but the DE perturbations are negligible, i.e.\ $\frac{\delta T^i_i}{\delta T^0_0}=0$. We solve numerically Eq.~(\ref{eq:potential}) for different values $w\in[-1,-0.7]$ and insert the resulting $\Psi(a)$ together with $H(a)$ in Eq.~\eqref{eq:nulltest}. The results are plotted in Fig.~\ref{fig:Oz_w}. We see that our null test is an efficient model-independent discriminator between \lcdm\ and a $w$CDM model. For $w=-0.9$ we already see deviations of the null test from unity and $w=-0.7$ would imply a deviation of around 4$\sigma$ from $\mathcal{O}(z)=1$.

In the second scenario, we account also for DE perturbations. More precisely, we fix the equation of state to $w=-0.95$, and we parameterize DE perturbations through the sound speed 
\begin{equation}
\small{
    \frac{\delta T^i_i}{\delta T^0_0}=\frac{\delta P_{\rm DE}}{\delta \rho_{\rm DE}+\delta \rho_{\rm DM}}=\frac{c^{2}_{s,{\rm DE}}\delta \rho_{\rm DE}}{\delta \rho_{\rm DE}+\delta \rho_{\rm DM}}=c^2_{s,{\rm eff}}}\, . \label{eq:cs2}
\end{equation}
In principle, the sound speed of DE $c_{s,{\rm DE}}^2$ can depend on both time and scale, i.e., $c^2_{s,{\rm DE}}=c_{s,{\rm DE}}^2(\tau,k)$. For instance, as mentioned in Ref.~\cite{Amendola:2015ksp}, the sound speed for a scalar field $\phi$  in the conformal Newtonian gauge at small scales is approximately  $c_{s,\phi}^2\simeq\frac{k^2}{4 a^2 m_\phi^2}$, where $m_\phi$ is the mass of the scalar field. Conversely, $c_s^2$ is equal to one only in the rest frame of the scalar field (see Chapter 11.2 of Ref.~\cite{Amendola:2015ksp} for a quick derivation). A similar situation occurs in $f(R)$ theories, as they effectively contain only a scalar degree of freedom \cite{Sawicki:2015zya}.\footnote{$f(R)$ theories can be interpreted as a non-minimally coupled scalar field in the Einstein frame.} Therefore, we expect the sound speed to be scale-dependent in modified gravity models when we are not in the rest frame of the equivalent DE fluid. Moreover, even in the case where $c_{s,{\rm DE}}^2$ is independent of scale, 
the effective sound speed introduced in Eq.~\eqref{eq:cs2} is expected to depend on scale 
since dark matter and DE perturbations do not share the same scale 
dependence. However, for simplicity here we consider an effective sound speed $c^2_{s,{\rm eff}}$ that 
is constant in time and in $k$. 

We solve numerically Eq.~(\ref{eq:potential}) for different values $c^2_{s,{\rm eff}}\in[5\cdot 10^{-7}, 10^{-5}]$. In this case, contrary to the previous scenario, the evolution of the potential depends on $k$ 
through the last term in Eq.~\eqref{eq:potential}, which does not vanish if $\delta T^i_i\neq 0$. As a consequence, $\mathcal{O}(z)$ acquires a scale 
dependence. In Fig.~\ref{fig:Oz_cs2} we show the resulting $\mathcal{O}(z)$ 
at the scale $k=300\,H_0\,\sim 0.1\,h\,\mathrm{Mpc}^{-1}$. The reason we select 
this specific value is that it corresponds to the largest value of $k$ we can choose without entering the non-linear regime. We compare the resulting $\mathcal{O}(z)$ with the \lcdm\ prediction and the $68.3\%$ confidence regions in the optimistic and pessimistic cases.
We see that our null test can discriminate between \lcdm\ and a DE model with perturbation. In the optimistic case, $c^2_{s,{\rm eff}}=1\cdot 10^{-6}$ already leads to deviations from unity at $1\sigma$. A value of 
$c^2_{s,{\rm eff}}=1\cdot10^{-5}$ on the other hand would imply a deviation of around 4$\sigma$ from $\mathcal{O}(z)=1$ for the pessimistic case and around 5$\sigma$ for the optimistic case. 

In Fig.~\ref{fig:Oz_cs2_1} we show a similar plot with the resulting $\mathcal{O}(z)$, but
at a larger scale (smaller $k$) of $k=150\,H_0\,\sim 0.05\,h \,\mathrm{Mpc}^{-1}$. In this case only sound speed values above $5\cdot 10^{-6}$ can be distinguished from $\Lambda$CDM. In practice, our method does not allow us to test for the scale-dependence of $\mathcal{O}(z)$ since the function $\hJ(z)$, which encodes the evolution of the Weyl potential, is assumed to be scale-independent. This means that in the case of a scale-dependent potential, the $\hJ$ measured from weak lensing data would be an averaged $\hJ$, over the range of available scales (and probably a poor fit to the data that require a scale-dependent $\hJ$). Given that lensing data contain information on the mildly non-linear regime, the sensitivity of the null test to such a scale-dependent scenario would probably be closer to Fig.~\ref{fig:Oz_cs2} than to Fig.~\ref{fig:Oz_cs2_1}.

\section{Conclusions  \label{sec:conclusions}}
For years, the \lcdm\ model has been a fundamental component of modern cosmology and the primary theoretical framework for current and future surveys. However, recent findings from DESI and other cosmological data suggest the presence of evolving DE. Due to the elusive nature of DE and the need to avoid model bias, we have presented a consistency test of the Weyl potential to test \lcdm\ with machine learning forecasts. In particular, we used the Noether's theorem approach in order to obtain a conserved quantity that can be written in terms of the Hubble rate $H(z)$ and the function $\hat{J}(z)$, which controls the evolution of the Weyl potential as a function of redshift.

To predict how effectively our new test, defined by Eq.~\eqref{eq:nulltest}, can constrain deviations from \lcdm\ at large scales, we generated mock datasets based on the specifications of DESI-like and LSST-like surveys, using the \lcdm\ model as the baseline cosmology across various profiles. This method enables us to quantify any deviations by utilizing forecasted mock data and plausible scenarios.

To reconstruct the $\mathcal{O}(z)$ statistic from Eq.~\eqref{eq:nulltest} using the mock data, we opted for a machine learning approach, specifically Genetic Algorithms (GA). This method enables us to obtain non-parametric and theory-agnostic reconstructions of the data, represented as $H(z)$ and $\hat{J}(z)$, which we can subsequently use to reconstruct $\mathcal{O}(z)$.
 
Using this approach, we find that the GA with the $\mathcal{O}(z)$ statistic accurately predicts the underlying fiducial cosmology across all redshifts covered by the data, as shown in Fig.~\ref{fig:Oz_w}. It can also confidently exclude several scenarios with a confidence level of ~$\gtrsim 4\sigma$ within the redshift range $z~[0.1-2]$.

\section*{Acknowledgements}
 DSB and CB acknowledge support from the Swiss National Science Foundation. CB also acknowledge funding from the European Research Council (ERC) under the European Union’s Horizon 2020 research and innovation program (Grant agreement No.~863929; project title ``Testing the law of gravity with novel large-scale structure observables"). RA and SN acknowledge support from the research project PID2021-123012NB-C43 and the Spanish Research Agency (Agencia Estatal de Investigaci\'on) through the Grant IFT Centro de Excelencia Severo Ochoa No CEX2020-001007-S, funded by MCIN/AEI/10.13039/501100011033.
\\

{\bf Numerical Analysis Files}: The GA codes used by the authors in the analysis of the paper can be found at  \href{https://github.com/RubenArjona}{https://github.com/RubenArjona}.\\

\bibliography{WL}

\end{document}